\documentclass[a4paper,11pt]{article}
\pdfoutput=1 

\usepackage{jheppub} 

\usepackage{fancyhdr}
\usepackage{graphicx}

\usepackage{braket}
\usepackage{dsfont}

\newcommand{\be}{\begin{equation}}
\newcommand{\ee}{\end{equation}}
\newcommand{\bea}{\begin{eqnarray}}
\newcommand{\nn}{\nonumber}
\newcommand{\eea}{\end{eqnarray}}

\newcommand{\Rmnum}[1]{\expandafter\@slowromancap\romannumeral #1@}

\newcommand{\cc}[2]{c\genfrac{[}{]}{0pt}{}{#1}{#2}}
\newcommand{\mb}[1]{\mathbf{#1}}
\newcommand{\omb}[1]{\overline{\mathbf{#1}}}
\newcommand{\yb}[0]{{\bar{y}}}

 \title{\LARGE Genetic Algorithms and the Search for Viable String Vacua}

\author[\diamondsuit]{Steven Abel}
\author[\dagger]{and John Rizos}

\affiliation[\diamondsuit]{IPPP,
Durham University, South Road, Durham, DH1 3LE, UK}
\affiliation[\dagger]{Physics Department, University of Ioannina, GR45110, Greece}

\abstract{\noindent Genetic Algorithms are introduced as a search method for finding string vacua 
with viable phenomenological properties. It is shown, by testing them against a class of 
Free Fermionic models, that they are orders of magnitude more efficient than a randomised 
search. As an example, three generation, exophobic, Pati-Salam models with a top Yukawa occur once in every 
$10^{10}$ models, and yet a Genetic Algorithm can find them after constructing only $10^5$ examples. Such non-deterministic search 
methods may be the only means to search for Standard Model string vacua with detailed phenomenological requirements.
}

\begin{document} 
\maketitle
\flushbottom

\section{Introduction}

Here is a physicist's estimate of the ``landscape of haemoglobin'':
it is composed of four chains of amino acids with either 141 or 146
in each chain; there are about 500 known amino acids; therefore the
haemoglobin molecule represents a solution out of a landscape of at
least $500^{574}$ possibilities. 

Such huge numbers lead one to wonder if the string landscape \cite{Douglas:2003um}, which 
is positively diminutive by comparison, might seem less daunting if it were approached using 
some of the many heuristic search tools based on natural processes, 
such as simulated annealing, tabu searches and
so on. In this paper we focus on the one most closely related to natural selection, namely Genetic
Algorithms (GA's) \cite{Holland,ga-books,david}. 

GA's (and indeed natural selection) work in systems where incremental
changes lead to incremental improvements. They are not very effective 
in ``needle-in-a-haystack''
situations where any solution other than the correct one is equally
disfavoured. String theory seems to be of the former variety where
a GA \emph{should} be applicable: the anecdotal evidence that supports
this is that, despite the fact that an exhaustive scan is clearly
out of the question, it has been possible in many different string
configurations (heterotic, intersecting branes, flux compactifications,
rational CFTs for example) to piece together models that closely resemble
the Standard Model. 

As such the task of finding a completely viable string vacuum is likely to be
what in computational complexity theory is called an NP-complete
problem (where NP refers to Non-deterministic Polynomial time); that
is a problem for which any given solution can be \emph{verified} in
a time that increases only polynomially with the difficulty, but where
finding a solution by a simple deterministic  search algorithm (such as exhaustive
scanning) rapidly becomes computationally infeasible. Indeed a similar point was made in 
ref.\cite{Denef:2006ad}, to which the reader is directed for precise definitions. NP-complete
problems are precisely where heuristic search methods become effective. 

The purpose of this paper is to demonstrate the efficacy of GA's in
finding desirable string vacuum solutions, by examining a small sub-class
of string theories, namely heterotic strings in the Free Fermionic formulation
\cite{Kawai:1986va,Antoniadis:1986rn,Kawai:1986ah}. We will show that they are (many orders of magnitude)
more efficient than a random search at finding string vacua with particular
desirable properties. This is especially evident when one applies
many phenomenological requirements and the search is multi-modal.
For example GA's do not confer much advantage if one is just searching
for say three generation models. However, in line with them being
effective on NP-complete problems, they come into their own when the
search is statistically very difficult (when for example only one
in $10^{7}$ models or fewer has the particular properties of interest).

Given the comments above, one thing we can conclude from the fact
that GA's work so well is that finding the SM in the string landscape
is precisely \emph{not }like looking for a needle in a haystack: the
landscape has structure and similar models have certain correlations.
We will describe exactly what these correlation are expected to
be, but because the number of possible models is so huge it is not possible for us
(even in this fairly restricted set of models) to check them explicitly.
Nevertheless in our view the fact that GA's work is evidence that they are there.

\subsection{Overview of GA's: a fake landscape of $10^{500}$}

Before getting to string theory, it is instructive to create a somewhat
artificial optimization problem that has a similarly large landscape
in order to introduce the GA technique and to make apparent its generic
advantages and also its limitations. Suppose that we wish to find
the supremum of some function $f(x,y)$ in the domain $x\in(0,10)$,
$y\in(0,10)$, without using calculus. One way do this is as follows:
consider writing out the possible coordinates 
\begin{eqnarray*}
x & = & a.bcdef...\\
y & = & g.hijkl...
\end{eqnarray*}
where $a,b,c...$ are digits between 0 and 9. In principle one could
scan over $x$ and $y$ by cycling through all possible strings of
digits $a,b,c...$ To find the supremum one simply evaluates $f(x,y)$
at each point, and chooses the largest value. Obviously this is a
very labour intensive way of solving the problem (which is why calculus
was invented): if the desired accuracy is high, to say 250 digits,
then one has to scan $10^{500}$ possible values of all the digits
to reproduce every possible value of $x$ and $y$. By wilful incompetence
the optimisation problem has been recast in a form that has a string-sized
landscape of possible solutions!!

The solution can be found with a GA as follows. We will describe a ``simple'' GA of the kind
developed by Holland \cite{Holland}. First introduce a
reasonably large population, $p$, of creatures defined by
their \emph{genotype}. (We discuss the optimal population size below.) This is simply a string of data 
for each one containing
all the defining properties -- often it is taken to be binary, but
in this case we will consider the strings of digits describing the
$x$ and $y$ coordinates to be two \emph{chromosomes}%
\footnote{The common terminology equates ``chromosome'' with ``genotype'',
but for stringy applications we think it may be convenient to
keep the distinction.%
} of length $250$. Each possible digit is referred to as an \emph{allele}, and 
each position as a \emph{locus}.
Initially the genotype of each member of the population
is chosen at random, so the population is effectively sprinkled at
random over the domain. Each genotype results in a physical characteristic,
the \emph{phenotype}, so called: in this case the phenotype can be
the function $f(x,y)$ evaluated at the point corresponding to the
genotype. 

The next step is to organise breeding, with pairs of creatures
being chosen to reproduce at random but with a weighting proportional
to their \emph{fitness}; in this case ``fitter'' creatures are obviously
those that are sitting higher on the landscape (``fitness landscape''
is coincidentally the usual GA terminology), so the fitness function
can simply be chosen to increase linearly with the height $f(x,y)$
in a way to be made precise below. The number of breeding pairs
is chosen to be $p$ so as to keep the population constant. Reproduction
consists of the following \emph{crossover} procedure: cut the chromosomes
of a breeding pair at the same two randomly chosen positions and swap
the middle section. (Usually but not always self reproduction is allowed.)
At the same time introduce some \emph{mutation}: reassign a tiny fraction
of the digits in the offspring (less than a percent usually) to arbitrary
random values\footnote{In this simple GA the entire population is killed off
and replaced at every generation (see \cite{ga-books}). There are many variations -- for
example the ``steady state'' GA continually produces new offspring,
and kills off the less fit members of the population for them to replace.
Likewise crossovers can be chosen to occur at one or many points on
the chromosomes, and selection can be organised differently. They all have in common though the three essential
ingredients of selection, crossover and mutation.}.

Remarkably, repeating this simplistic procedure over a number of generations
results in a population that gathers with increasing accuracy around
the desired global maximum value of the function. Moreover the three ingredients 
of selection, crossover and mutation are crucial\footnote{Note crossover and mutation can be either/or; they should be present in the population but do not have to occur simultaneously in the same individuals.}. If done correctly
(see below) one can obtain a solution to any desired precision. 

It is worth noting some advantages over other techniques. First the
function $f$ can have many maxima, and yet the procedure can still
find the global one: the algorithm effectively samples the whole fitness
landscape. Indeed $f$ does not even need to be differentiable, a
fact that strongly suggests the technique could be powerful in the
string context, where getting from vacuum to vacuum often
involves topology changing transitions. In addition, the computational
difficulty appears to rise roughly linearly with the length of the genotype even
though the size of the fitness landscape is increasing exponentially. 
Finally, the process is very robust. It doesn't matter for example
if we choose to flatten all the chromosomes of each creature into
one long string of data and perform a single crossover for the entire
genotype, or if we perform crossovers on the chromosomes individually. 

\begin{figure}
\noindent \centering{}\includegraphics[scale=0.6]{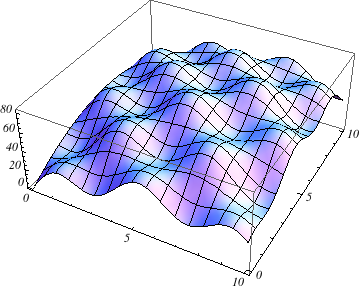}
\caption{\em The ``mogul-field'' function.\label{fig:The-test-function} }
\end{figure}

Many of these properties can be understood (at least intuitively) in terms of \emph{schemata}
and the \emph{schema theorem} which was introduced by Holland \cite{Holland} and which 
we will describe shortly. But
before we do so, it is worth seeing the procedure at work on a particular
function. Consider finding the maximum of
\begin{equation}
f(x,y)=12\left(\cos\frac{3y}{2}\,\sin\frac{3x}{2}+x+y\right)-x^{2}-y^{2}.
\end{equation}
This ``mogul-field'' function, shown in figure~\ref{fig:The-test-function},
is clearly a hard function for the usual hill-climbing algorithms
to maximize. 

As mentioned above, the simplest convention for choosing breeding pairs, and the one we shall use here, 
is that they are weighted linearly with $f(x,y)$ (roulette wheel selection): that
is they are selected with a probability given by a fitness function
for the $i$'th creature of the form 
\begin{eqnarray}
p_{i} & = & \frac{af(x_{i},y_{i})+b}{\sum_{i}(af(x_{i},y_{i})+b)},\\
\nonumber 
\end{eqnarray}
where $a$ and $b$ are constants that need to be adjusted each generation.
If the weighting is chosen with $a$ too large, one finds that fitter
creatures swamp the distribution very early, and there is premature
convergence to the wrong solution. Once the population is all gathered
around the wrong peak, the advantage of selection is lost:
one has to hope for an advantageous random mutation, so in a sense
the process has temporarily become no more efficient than a Monte-Carlo
procedure (although once the population has ``Monte-Carlo'd'' out
of the wrong solution, the process reverts to the preferred GA behaviour).
There are various sophisticated techniques to counteract premature convergence, such
as introducing a fitness penalty for crowding, or fitness sharing.
We will not need to explore these here, but in the event of such stagnation
will instead resort to momentarily enhancing the mutation rate. Why
this works will become apparent when we come to discuss the schema
theorem below, but generally it is already clear that selection and mutation are playing 
complementary roles; selection drives the system to nearby maxima, while mutation 
drives one away. 

Conversely once the population is gathered around the correct global peak
one can increase $a$ and dial down mutation in order to distinguish 
between creatures that all have similar heights, and gain accuracy. 
The convention is to choose $a$ and $b$ such that a creature of average fitness
breeds once, while the fittest creature breeds twice on average. If
the latter multiple is $\alpha$, then the fitness function becomes
\[
p_{i}=\frac{1}{p}\frac{(\alpha-1)\left(f_{i}-\bar{f}\right)+\left(f_{max}-\bar{f}\right)}{f_{max}-\bar{f}},
\]
with the average fitness $\bar{f}$ and maximum fitness $f_{max}$
being re-evaluated for every generation. Note that the probability
is invariant under rescaling and shifting of the function $f(x,y)\rightarrow\beta f(x,y)+\gamma$.
In our analysis we found that a higher value, $\alpha\approx3$, gave
a slightly better convergence rate. In this particular example (but
not actually in the string theory problems we shall discuss) one final
adjustment can make convergence more rapid: De Jong's Elitist Selection
Scheme involves copying the fittest member
of the population across to the next generation unchanged. 

The evolution of a population of 60 individuals is shown in figure
\ref{fig:The-test-function-1}. The initial population coalesces around
the maxima after only a few generations. The lower maxima are then
disfavoured until the population is all gathered around the solution
(which for the record is at $x=6.26347798$ and $y=7.23832285$).
Note that around generation 10 the population seems to be favouring
the wrong maximum but then corrects itself after a few generations.

There are two aspects that limit the final accuracy%
\footnote{To achieve this accuracy the computing time required in Python coded
on a conventional PC is a few seconds. Achieving the same accuracy
doing a scan would take about a millenium.%
}. The first is simply the precision to which the function $f(x,y)$
can be evaluated when the algorithm reaches a plateau. The second is the population 
size required for a meaningful search, which increases with the genotype length. As discussed 
in ref.\cite{ga-books} one useful criterion to adopt is that every point in the search space could in principle be reached by cross-over only, which requires that in the initial population every 
allele should occur at least once at each locus with say $P_*=0.999\%$ probability. For a decimal ``alphabet'' this 
exacts a relatively high price in terms of population size as the length increases, whereas for a binary one (assuming initially randomised genotypes) the minimum population required is
$$
p\approx [ 1+\log(-\ell /\log P_*)/\log 2 ]
$$
which increases very slowly. This is one reason it is considered more efficient to express the genotype in a binary 
form. (e.g. using a binary genotype an accuracy of $10^{500}$ would require a genotype of length $\ell  = 1661$, and a \emph{minimum} population of only $p=22$).

Nevertheless it is important to recall that the chromosomes' length is 250, so
the process really is sampling the whole $10^{500}$ landscape even if the 
population is only 60. As already mentioned, the process can cope with complicated
and non-differentiable landscapes: for example figure \ref{fig:The-test-function-1-1}
shows the 500'th generation for $f(x,y)=12\left(3\cos(50y)\,\sin(50x)+x+y\right)-x^{2}-y^{2}.$
This function is extremely ``choppy'', having $\mathcal{O}(50^{2})$
relatively high maxima and minima in the domain. GA's are impeded by more rugged landscapes but 
can still operate. This illustrates another feature of
GA's which is that because it begins widely spread over the domain,
the population initially averages over rapid local fluctuations and
responds to the longer wavelength features. Despite this robustness,
it is clear that even though the fitness landscape does not need to
be differentiable, it should certainly have structure. For example
if we had a ``needle in a haystack'' landscape consisting of $10^{500}-1$ squares in which
$f=0$ and one square with $f=1$, then a GA would be no better than
a random scan at landing on this one square. 

\noindent \begin{center}
\begin{figure}
\begin{tabular}{cc}
\includegraphics[scale=0.35]{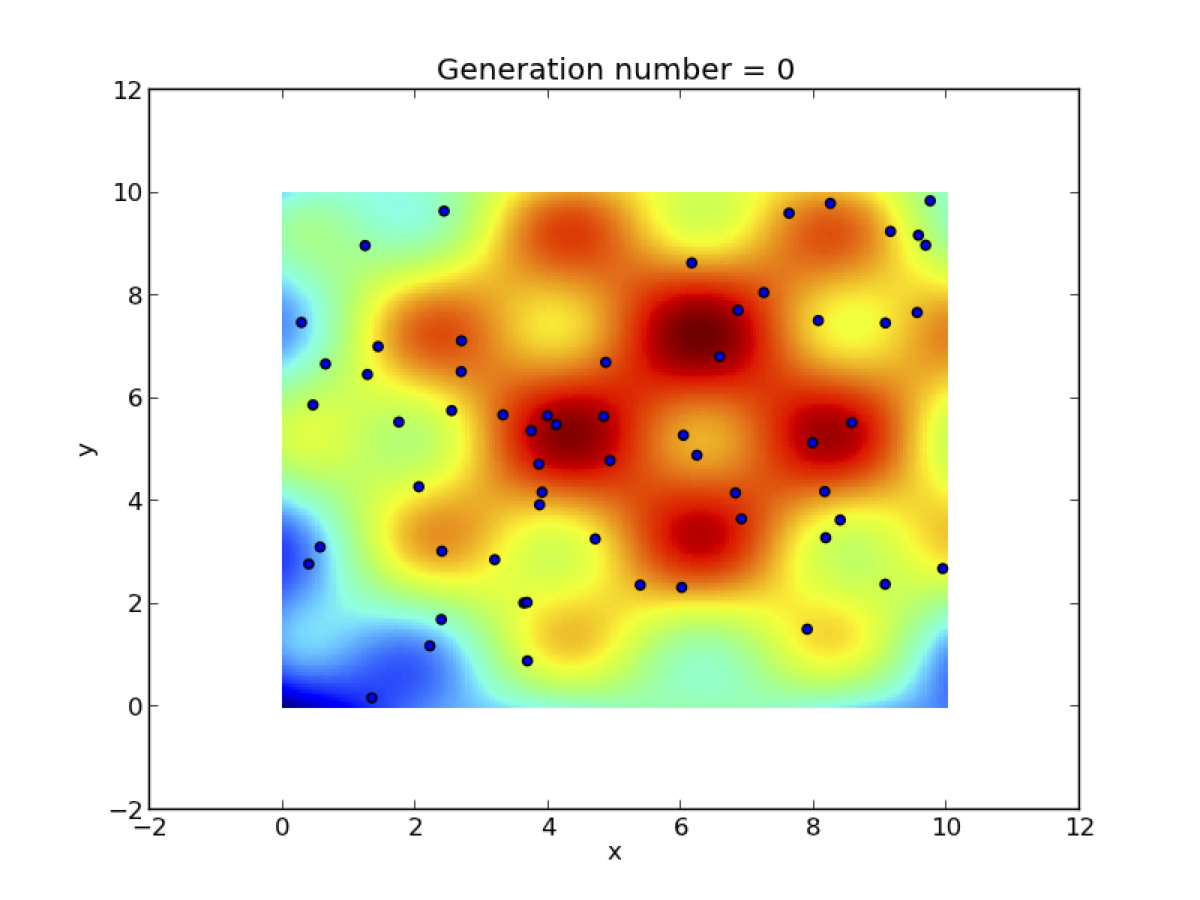} & \includegraphics[scale=0.35]{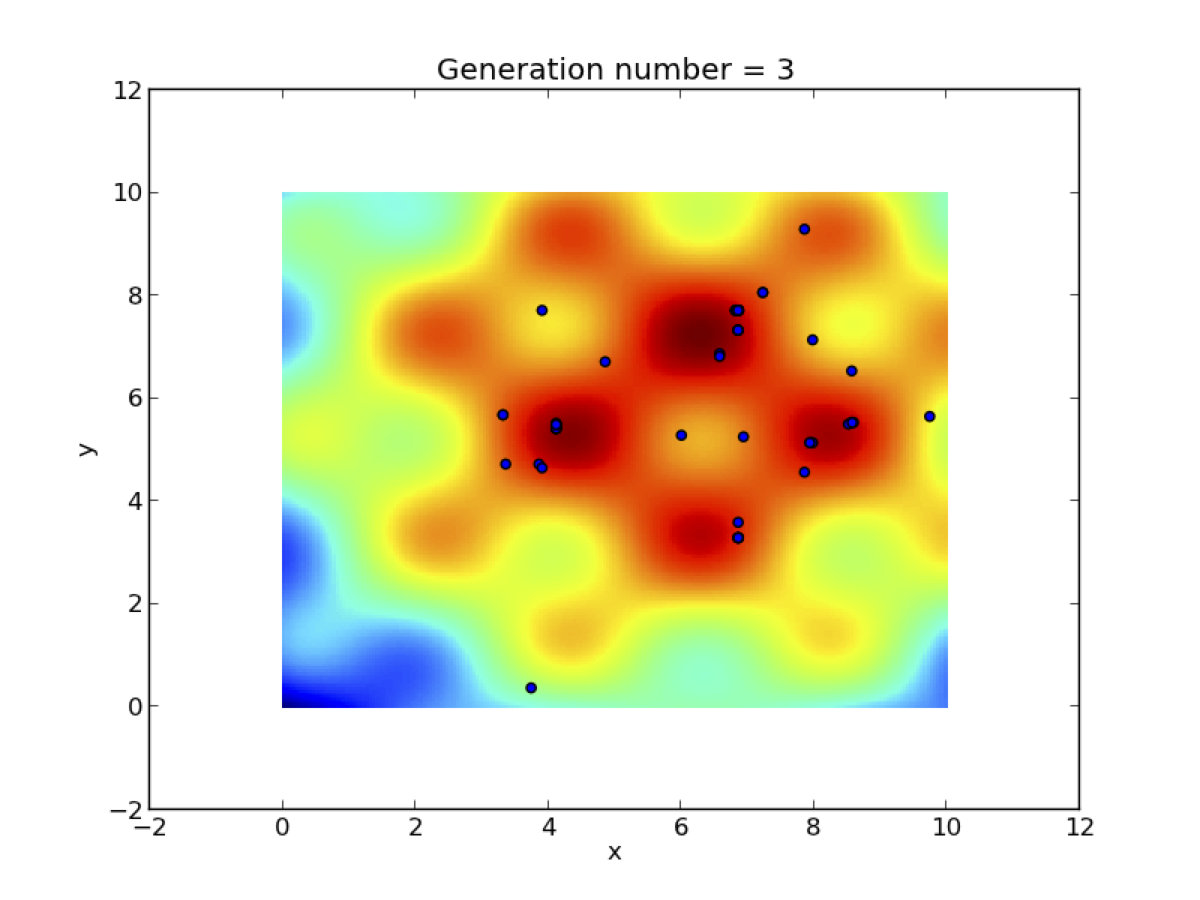}\tabularnewline
\includegraphics[scale=0.35]{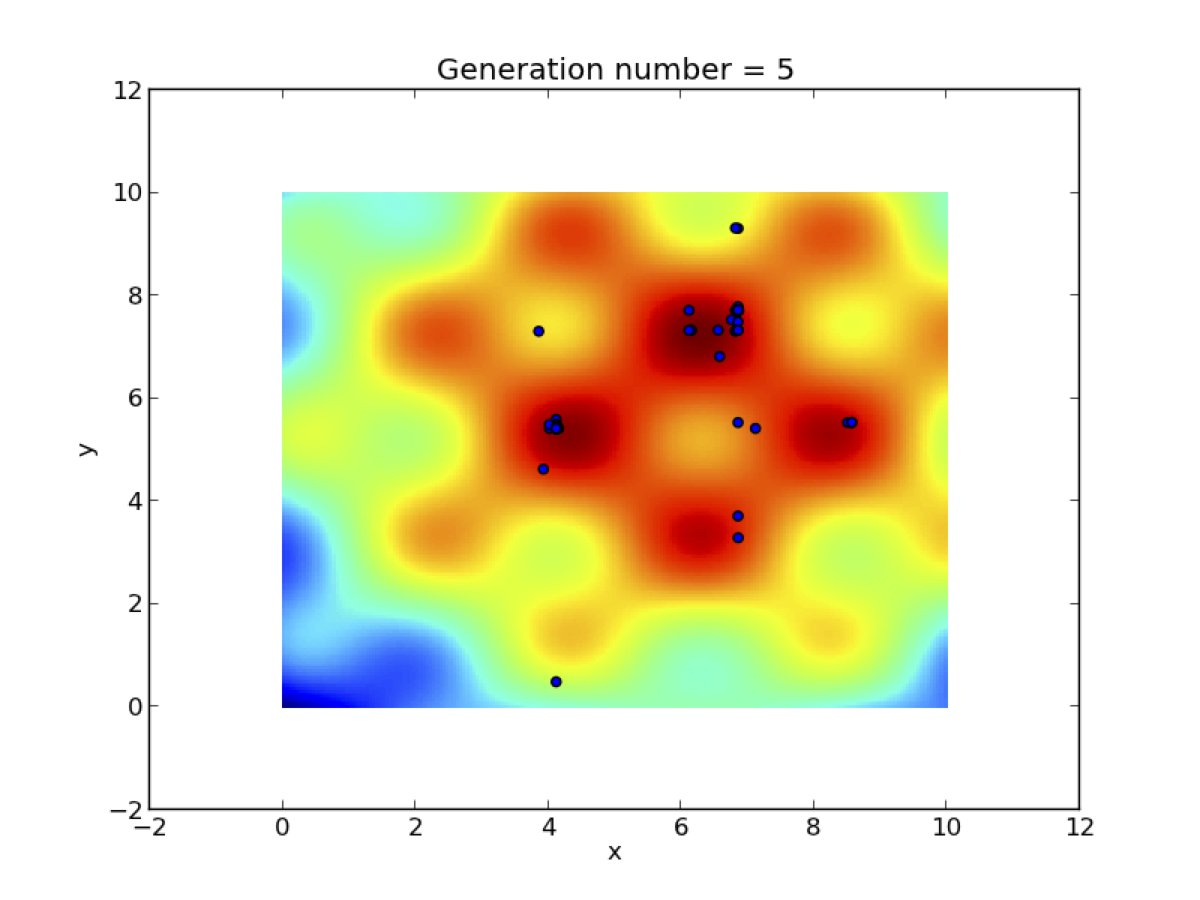} & \includegraphics[scale=0.35]{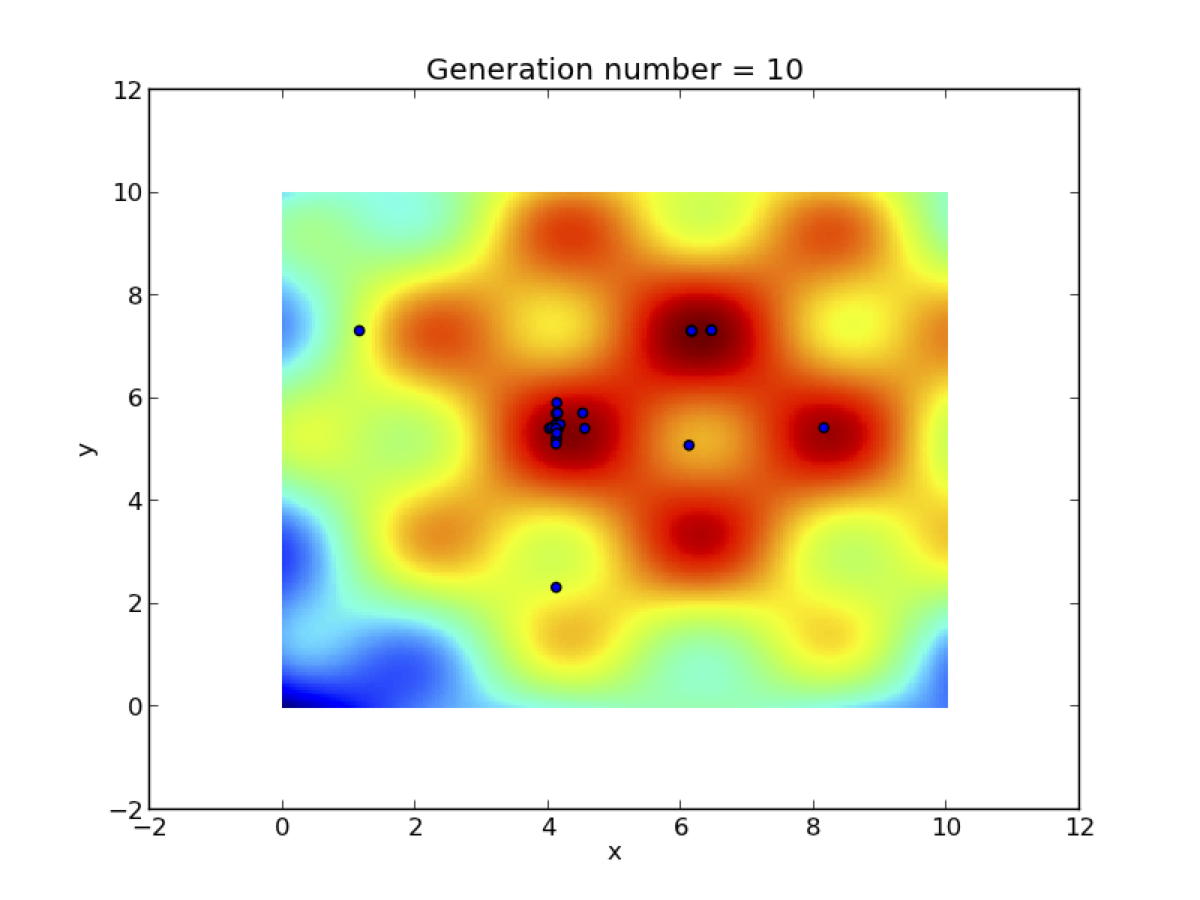}\tabularnewline
\includegraphics[scale=0.35]{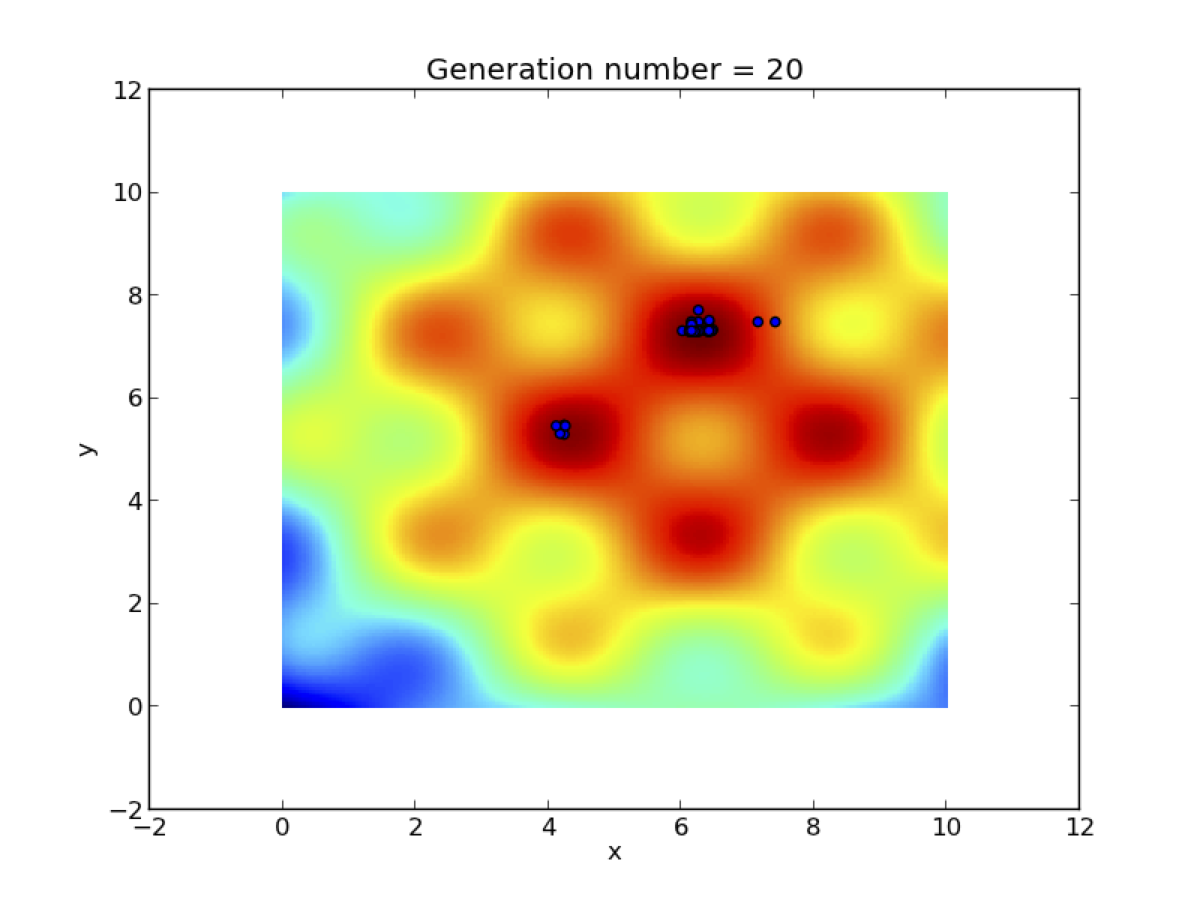} & \includegraphics[scale=0.35]{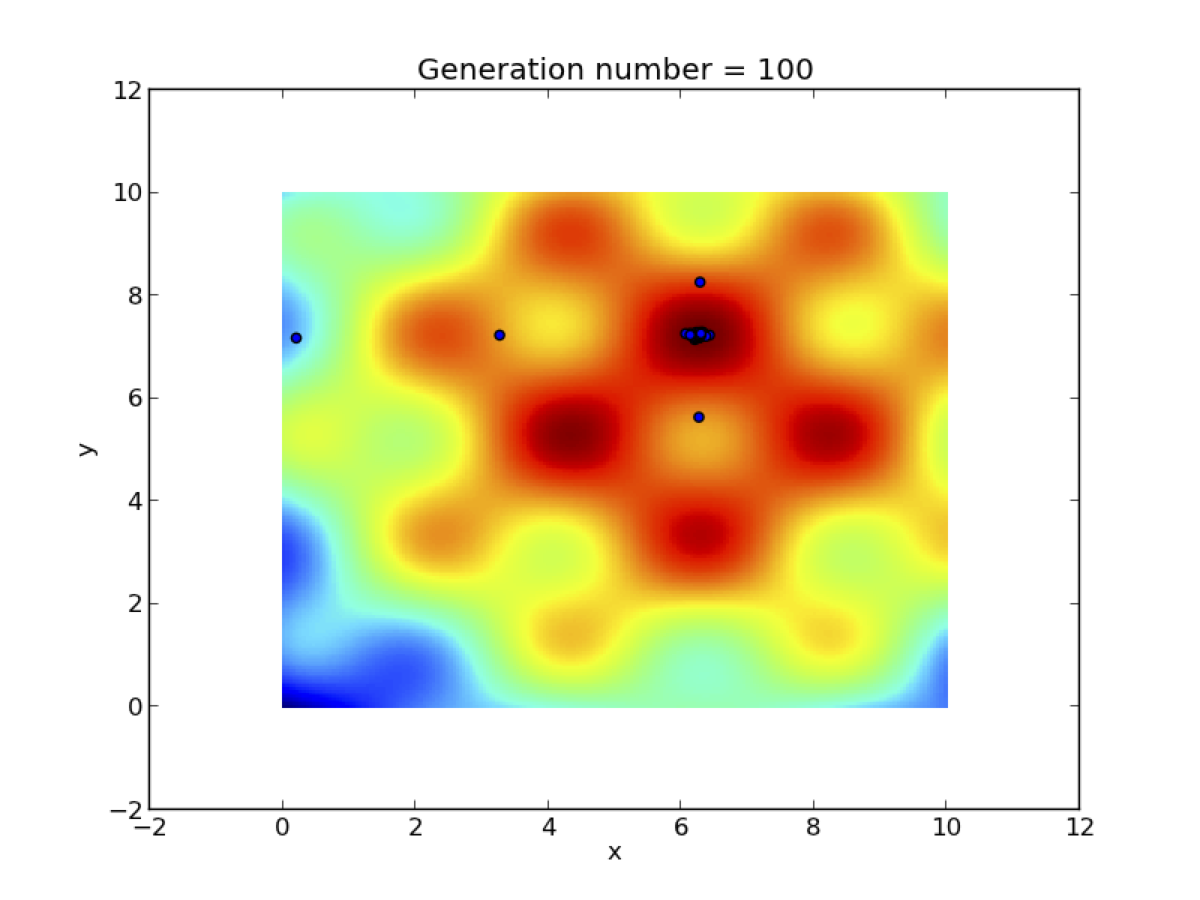}\tabularnewline
\end{tabular}\caption{\em Evolution of a population of 60 individuals in a landscape of $10^{500}$.\label{fig:The-test-function-1} }
\end{figure}
\begin{figure}
\noindent \centering{}\includegraphics[scale=0.5]{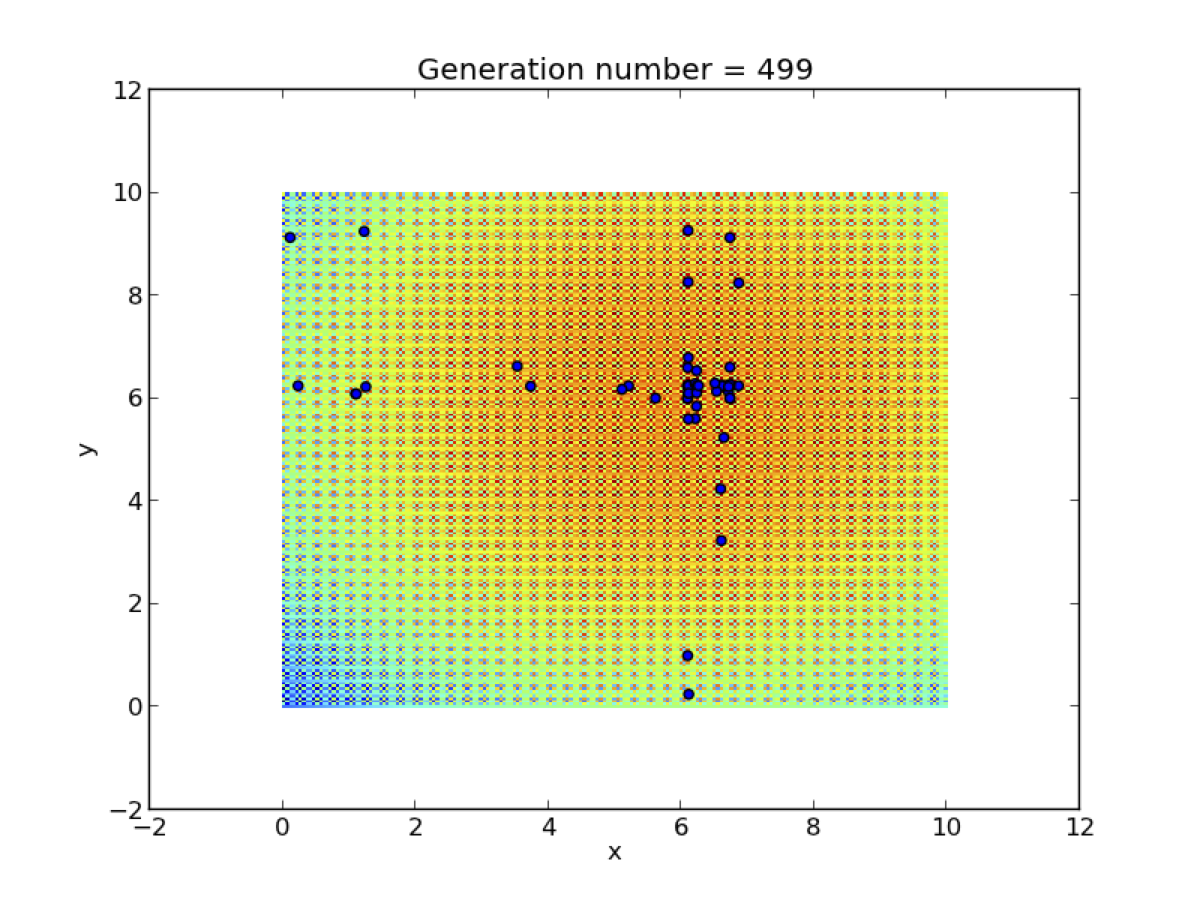}\caption{\em Evolved population of 60 individuals in an ``almost discontinuous''
extremely choppy landscape.\label{fig:The-test-function-1-1} }
\end{figure}

\par\end{center}

\subsection{The schema theorem}

As we mentioned, the \emph{schema theorem} was introduced by Holland
as a way to formalise the remarkable properties of GA's \cite{Holland}. It has been 
criticised by many authors but nevertheless, while nowadays it is not thought to give a complete understanding, it 
does present useful ideas that at least partially explain how GA's work. 
 
A schema is a representation of some crucial set of digits that is supposed
to confer some favourable characteristic, and in this instance might
look something like $S=3***4*6$. This example has 3 entries that
we are interested in (hence we say it is order $3$) and 4 entries
that we do not care about, which are labelled with a wildcard $*$.
It also has a defining length $d(S)=7$, and we will call its order,
$o(S)$.

Holland argued that schemata are important because selection favours
the propagation of shorter strings of data: small subsections of the
genome that confer fitness dominate first and, once they are shared
by the majority of the population, crossover does not affect them.
Indeed this can be observed directly in our previous example: the
population tends to spread along the $x$ and $y$ directions from
the solution because in this example the approximately correct $x$
and $y$ values correspond to only the first few entries of the $x$
and $y$ chromosomes, which tend to persist even though the entire
genome may be disturbed by crossover.

This can be formalised as follows. Suppose that mutation has just
produced in the population a favourable schema, $S$. Let $n(S,t)$
be the total number in the population containing it at time $t$.
We can define the average fitness of all members of the population
containing $S$, as $f_{S}(t)=\sum_{i\in S}f_{i}/n(S,t)$, which is
higher than the average fitness of the population as a whole, $\bar{f}$.
Assuming that selection is proportional to fitness, $f(t)$, then
the expected number of offspring containing $S$ is $\sum_{i\in S}f_{i}/\bar{f}$.
Neglecting crossover and mutation this would be the expectation of
$n(S,t+1)$; let us rewrite it as 
\begin{equation}
n(S,t+1)=n(S,t)\frac{f_{S}(t)}{\bar{f}}.
\end{equation}
With simple probabilistic arguments one can incorporate the effect
of a single-point crossover destroying $S$, and mutations at a rate
$p_{m}$ per digit to find a lower bound 
\begin{equation}
n(S,t+1)\geq n(S,t)\frac{f_{S}(t)}{\bar{f}}\left(1-\frac{d(S)}{l-1}\right)\left(1-p_{m}\right)^{o(S)},
\end{equation}
where $l$ is the length of the entire genome. It is a lower bound
because obviously if both parents have the schema $S$ then it cannot
be destroyed by crossover. When very few members of the population
have the schema then it is an approximate equality, whereas if most
of the population have the schema then
\begin{equation}
n(S,t+1)\approx n(S,t)\frac{f_{S}(t)}{\bar{f}}\left(1-p_{m}\right)^{o(S)}.
\end{equation}

These relations capture the observed growth of favourable schema.
If our new schema gives a greatly improved fitness $\bar{f}(1+\Delta)$
then at first we can neglect the $n(S)$ contributions in $\bar{f}$
and also the crossover and mutation, so that the subsequent evolution
is given by 
\begin{equation}
\frac{dn(S,t)}{dt}\approx n(S,t)\Delta,
\end{equation}
and the occurrence of schema $S$ grows exponentially%
\footnote{We should remark that we are presenting the original and most simplistic
interpretation which will suffice for a qualitative understanding
of the behaviour. There has been debate about the proper interpretation
and in particular about how long exponential growth continues until
it is affected by the change in the make-up of the population (see Reeves and Rowe in \cite{ga-books}).
In particular it should be borne in mind that the optimal population is typically 50-100 individuals, so that 
the exponential growth phase can only be a short lived idealisation.
}. This continues until the schema begins to saturate the population
at which point the average fitness $\bar{f}$ becomes dominated by
it. The system begins to approach the situation where every member
of the population contains $S$, and hence we can write $\bar{f}=f_{S}(1-\delta)$
where $\delta$ is approaching zero. Clearly $\delta$ has an equilibrium
value, 
\begin{equation}
\delta=o(S)\, p_{m},
\end{equation}
so that $S$ is not quite able to saturate the entire population.

Thus we see the trade-off of GA's which is that the entire population
can never quite reach the perfect solution, although some members
may do. On the plus side however, it could be that schema $S$ represents
a local optimum but that our desired solution has an even better schema
in mind. In that case, as we mentioned above, as well as producing new schemata,
mutation plays a second role in that it prevents the GA stagnating.
This is why increasing the mutation rate temporarily is sufficient
to free the system if it becomes trapped at a local extremum. (Note however
-- and we shall see this in practice when we come to apply GA's to
string vacua -- that the most efficient procedure by far is to mutate
roughly 10-20\% of the genotype rather than to scramble them completely
which would lose all the beneficial schemata that had been acquired
up to that point.) In this manner we expect the algorithm to proceed
by bursts of rapid improvement followed by periods of stasis towards
the desired global solutions.

\section{The problem: viable Free Fermionic  Pati--Salam vacua}

We now present the stringy problem that we will consider for this study, namely finding phenomenologically viable Pati-Salam models in the 
Free Fermionic Formulation of the heterotic superstring \cite{Kawai:1986va,Antoniadis:1986rn,Kawai:1986ah}.

Before we describe the formalism in detail, let us briefly comment further on the relation of our approach to the landscape programme. It has been known for a long time that these and similar models lead to a huge number of possible vacua. For example 
\cite{Lerche:1986cx} estimated $10^{1500}$ vacua in the closely related covariant lattice approach, far in excess even of the later flux vacua estimate in \cite{Douglas:2003um}. The approach advocated in \cite{Douglas:2003um} and related papers (see \cite{Douglas:2012bu}
for a recent review) was to determine correlations between physical characteristics. Alternatively one can count the multiplicities of string vacua and regard the characteristics that occur frequently as being more natural. 

Completely general computer-based searches were used to consider correlations for the Free Fermionic vacua in ref.\cite{ffvs}. 
However, there are limitations to these and similar approaches, due to the space of 
models being so large, and due to the time-consuming computation of the spectrum in every step of the search procedure. Importantly this leads to inevitable restrictions as to what statistical correlations can and cannot reliably be established, as discussed in ref.\cite{Dienes:2006ca}. 

As we shall see, in performing a GA study one is also effectively studying correlations, but very different ones from those that were explored in the landscape programme. In the language of GA's the difference is that essentially the latter explored phenotype-phenotype correlations, whereas the frequencies occurring in GA studies are more sensitive to genotype-phenotype correlations, in a way that will be made precise below.

Now to the formulation, in which consistent models are defined 
in terms of a set of basis vectors $$\{v_1,v_2,\dots,v_n\}$$ and a set of phases $$\cc{v_i}{v_j},i,j=1,\dots,n$$ associated  with 
generalised GSO projections (GGSO). The basis vectors and the GGSO phases are subject to constraints that guarantee modular invariance of the one loop partition function. The elements of the basis vectors are related to the parallel transportation properties of the fermionised world-sheet degrees of freedom along the non-contractable torus loops. This yields models directly in four space-time dimensions with internal coordinates fixed at the fermionic point.


In order to test GA's against a set of models where a comprehensive scan is feasible for comparison, 
we will use the efficient, albeit not general, hybrid method that was formulated in  
\cite{cformalism} for the case of $Z_2\times Z_2$ vacua. It is based on a fixed set of basis vectors,  singling out models with $SO(10)$ gauge symmetry that play the role of the ``observable'' gauge group, and a variable set of GSO projection coefficients. This set-up permits the derivation of analytic results for several characteristics of the models,
including the number of fermion generations, the number of additional vector-like fermion/Higgs multiplets, the number of SM breaking Higgs doublets and the number of exotic multiplets. Integrating these analytical formulae into a computer programme we can achieve a much higher scan speed of the order of $10^4$ models per second for a single CPU core. 

The first class of models studied in this manner were Pati--Salam vacua \cite{pshm}.
The Pati-Salam GUT model possesses  $SU(4)\times{SU(2)}_L\times{SU(2)}_R \supset SO(10)$ gauge symmetry \cite{ps}. In the supersymmetric version \cite{pss}  SM quarks and leptons reside in
$SO(10)$ spinorials, $\mb{16}=\left(\mb{4},\mb{2},\mb{1}\right)+\left(\omb{4},\mb{1},\mb{2}\right)$,
\begin{align}
 \left({\mathbf4},{\mathbf2},{\mathbf1}\right)&=Q\left({\mathbf{3}},{\mathbf{2}},+\frac{1}{6}\right) + L\left({\mathbf{1}},{\mathbf{2}},-\frac{1}{2}\right)
 \\
\left({\overline{\mathbf4}},{\mathbf1},{\mathbf2}\right)&=u^c\left(\overline{\mathbf{3}},{\mathbf{1}},-\frac{2}{3}\right)+
d^c\left(\bar{\mathbf{3}},{\mathbf{1}},+\frac{1}{3}\right)+e^c\left({\mathbf{1}},{\mathbf{1}},1\right)+
\nu^c\left({\mathbf{1}},{\mathbf{1}},0\right)\,,
\end{align}
while SM Higgs doublets  together with additional triplets are accommodated in $SO(10)$ vectorials, ${\mathbf{10}}=\left(\mb{1},\mb{2},\mb{2}\right)+\left(\mb{6},\mb{1},\mb{1}\right)$,
\begin{align}
 \left({\mathbf1},{\mathbf2},{\mathbf2}\right)&=H_u\left({\mathbf{1}},{\mathbf{2}},+\frac{1}{2}\right)+
 H_d\left({\mathbf{1}},{\mathbf{2}},-\frac{1}{2}\right)\\
\left({{\mathbf6}},{\mathbf1},{\mathbf1}\right)&=
\delta\left({\mathbf3},{\mathbf1},-\frac{1}{3}\right) + \delta^c\left(\overline{\mathbf3},{\mathbf1},+\frac{1}{3}\right)\,.
\end{align}
In addition breaking of the PS symmetry requires the presence of at least one pair
of  Higgs fields in the representations
\begin{align}
 \left({\mathbf4},{\mathbf1},{\mathbf2}\right)_H&=
 u_H\left({\mathbf{3}},{\mathbf{1}},+\frac{2}{3}\right)+
  d_H\left({\mathbf{3}},{\mathbf{1}},-\frac{1}{3}\right)+
  e_H\left({\mathbf{1}},{\mathbf{1}},-1\right)+
  \nu_H\left({\mathbf{1}},{\mathbf{1}},0\right)\\
 \left({\overline{\mathbf4}},{\mathbf1},{\mathbf2}\right)_H&=
 u^c_H\left(\overline{\mathbf{3}},{\mathbf{1}},-\frac{2}{3}\right)+
 d^c_H\left(\bar{\mathbf{3}},{\mathbf{1}},+\frac{1}{3}\right)+
 e^c_H\left({\mathbf{1}},{\mathbf{1}},+1\right)+
 \nu^c_H\left({\mathbf{1}},{\mathbf{1}},0\right)\,,
\end{align}
which belong to an $SO(10)$ spinorial/antispinorial pair. In the minimal scenario, 
of one PS Higgs pair, the GUT Higgs mechanism removes $u_H, u^c_H, e_H, e^c_H$ and one combination of $\nu_H,\nu^c_H$ from the 
low energy spectrum. The leftover Higgs triplets $d_H,d^c_H$ may obtain heavy masses combined with the additional triplets in $\left(\mb{6},\mb{1},\mb{1}\right)$ through the couplings  
\begin{align}
\left(\mb{4},\mb{1},\mb{2}\right)_H^2\,\left(\mb{6},\mb{1},\mb{1}\right)+
\left(\omb{4},\mb{1},\mb{2}\right)_H^2\,\left(\mb{6},\mb{1},\mb{1}\right)\,=
d_H\,\delta^c\,\left<\nu_H\right>+d^c_H\,\delta\,\left<\nu^c_H\right>+\dots
\end{align}
In general, the spectrum of the model also includes a number of singlets  $\phi\left(\mb{1},\mb{1},\mb{1}\right)$ that may couple to the additional triplet and Higgs
fields. Some of these singlets are likely to develop vevs and provide mass terms for the non chiral fields of the spectrum.

Fermion masses arise from the coupling
\begin{align}
\left({\mathbf{4}},{\mathbf{2}},{\mathbf{1}}\right)\,\left(\bar{\mathbf{4}},{\mathbf{1}},{\mathbf{2}}\right)\,\left({\mathbf{1}},{\mathbf{2}},{\mathbf{2}}\right)=Q\,u^c\,H_u+Q\,d^c\,H_d+L\,e^c\,H_d+L\,\nu^c\,H_u \,.
\end{align}
Mixing of neutrinos with some of the (heavy) singlet states is required in order to attain realistic neutrino masses.

Summarising, the spectrum of a supersymmetry PS GUT model includes \vspace{0.3cm}\\
\begin{tabular}{ll}
 $n_g$ & $\times$ fermion generations in $\left(\mb{4},\mb{2},\mb{1}\right)+\left(\omb{4},\mb{1},\mb{2}\right)$\nonumber \\
$k_R$ &  $\times$ PS Higgs pairs transforming as
$\left({\mathbf4},{\mathbf1},{\mathbf2}\right)+\left(\omb{4},{\mathbf1},{\mathbf2}\right)$ \\
$n_h$ &  $\times$ SM Higgs doublet pairs in  
$\left({\mathbf1},{\mathbf2},{\mathbf2}\right)$\\
$n_6$ & $\times$  additional triplet pairs in $\left({\mathbf6},{\mathbf1},{\mathbf1}\right)$,
\end{tabular} \vspace{0.3cm}\\
together with a number of singlet fields. Obviously for a realistic model $n_g=3$ and $k_R\ge1$, $n_h\ge1$ , $ n_6\ge 1$, with equalities corresponding to the minimal models. 
In terms of PS gauge group representations there is an overlap between Higgs and generation fields, as they both contain  $\left({\omb{4}},{\mb{2}},{\mb{1}}\right)$ representations. Moreover, the spectrum may include a number, $k_L$, of additional (left) vector-like fields, 
$\left(\mb{4},\mb{2},\mb{1}\right)+\left(\omb{4},\mb{2},\mb{1}\right)$.
Denoting by $n_{42}, n_{4\overline{2}},n_{\overline{4}2}$ and $n_{\overline{4}\overline{2}}$ the number of $\left(\mb{4},\mb{2},\mb{1}\right)$, $\left(\mb{4},\mb{1},\mb{2}\right)$, $\left(\omb{4},\mb{2},\mb{1}\right)$ and  $\left(\omb{4},\mb{1},\mb{2}\right)$ fields respectively we obtain the following relations
\begin{align}
n_{{4}2}-n_{\overline{4}2}=n_{\overline{4}\overline{2}}-n_{{4}\overline{2}}=n_g\label{gc}\\
n_{\overline{4}2}=k_L\ ,\ n_{4\overline{2}}=k_R\,.
\end{align}
Equation \eqref{gc} can be regarded as a constraint  that guarantees the integrity of fermion generations. As a result, putting aside singlet states, as far as the spectrum is concerned  a PS model can be characterised by five integers namely $n_g,k_L,k_R,n_h,n_6$. A minimal model has $n_g=3, k_L=0, k_R=1,n_h=1,n_6=1$, however any model with $n_g=3, k_L\ge0, k_R\ge1,n_h\ge1,n_6\ge1$ is in principle phenomenologically acceptable at this level of analysis.

Semi-realistic supersymmetric Pati--Salam models can be relatively easily constructed in 
the Free Fermionic Formulation \cite{pss,spss}. They require exclusively 
periodic-antiperiodic boundary conditions on the world-sheet fermions and the corresponding GGSO phases can be written in terms of binary integers, $\cc{v_i}{v_j}=e^{i\,\pi\,c_{ij}}, c_{ij}=\{0,1\}$.
Following \cite{pshm} we can systematically study a class of PS models generated by the basis vector set $B=\left\{v_1,v_2,\dots,v_{13}\right\}$, where
\begin{align}
v_1=\mathds{1}&=\left\{\psi^\mu,\
\chi^{1,\dots,6},y^{1,\dots,6},\omega^{1,\dots,6}|\yb^{1,\dots,6},
\bar{\omega}^{1,\dots,6},\bar{\eta}^{1,2,3},
\bar{\psi}^{1,\dots,5},\bar{\phi}^{1,\dots,8}\right\}\nonumber\\
v_2=S&=\left\{\psi^\mu,\chi^{1,\dots,6}\right\}\nn\\
v_{2+i}=e_i&=\left\{y^{i},\omega^{i}|\bar{y}^i,\bar{\omega}^i\right\}, \ i=1,\dots,6\nn\\
v_{9}=b_1&=\left\{\chi^{34},\chi^{56},y^{34},y^{56}|{\yb}^{34},{\yb}^{56},
\bar{\eta}^1,\bar{\psi}^{1,\dots,5}\right\}\label{psbasis}\\
v_{10}=b_2&=\left\{\chi^{12},\chi^{56},y^{12},y^{56}|\bar{y}^{12},\bar{y}^{56},
\bar{\eta}^2,\bar{\psi}^{1,\dots,5}\right\}\nn\\
v_{11}=z_1&=\left\{\bar{\phi}^{1,\dots,4}\right\}\nn\\
v_{12}=z_2&=\left\{\bar{\phi}^{5,\dots,8}\right\}\nn\\
v_{13}=\alpha&=\left\{\bar{\psi}^{45},\yb^{1,2}\right\}\ .
\end{align}
Here we denote the fermionised world-sheet coordinates  as follows:  $\psi^\mu$ , $\chi^I$, ${I=1,\dots,6}$ are the superparteners of the 10-dimensional left-moving coordinates, $y^I,\omega^I/\bar{y}^I,\bar{\omega}^I$, ${I=1,\dots,6}$ stand for six internal  left/right coordinates, and $\bar{\psi}^A, A=1,\dots,5$, $\bar{\eta}^\alpha,\alpha=1,2,3$, $\bar{\phi}^k,k=1,\dots,8$ are the additional right-moving  complex fermions. 
We have adopted the traditional (ABK) notation where the fields included in a basis vector set are anti-periodic while the rest are periodic. 

The associated generalised GSO coefficients are not fixed but they are constrained by modular invariance. Consequently only the $\cc{v_i}{v_j}, i>j$ are independent. Moreover, the requirements of space-time supersymmetry fix some of these coefficients while some others are set by convention. Altogether, only 51 independent GGSO phases are relevant to the ``observable'' PS spectrum. These can be parametrised in terms of $\ell_i=\{0,1\},i=1,\dots,51\,$, as follows
\begin{align}
c_{ij}=
\bordermatrix{
      &\mathds{1}  & S&e_1&e_2&e_3&e_4&e_5&e_6&b_1&b_2&z_1&z_2&\alpha\cr
\mathds{1} & 1	& 1& 1& 1& 1& 1& 1& 1& 1& 1& 1& 1& 1\cr
S  	& 1	& 1& 1& 1& 1& 1& 1& 1& 1& 1& 1& 1& 1\cr
e_1	& 1	& 1& 0& \ell_{26}& \ell_{27}& \ell_{28}& \ell_{29}& \ell_{30}& \ell_{6}&0&\ell_{14}& \ell_{20}&\ell_{41}\cr
e_2	& 1	& 1& \ell_{26}& 0& \ell_{31}& \ell_{32}& \ell_{33}& \ell_{34}& \ell_7& 0& \ell_{15}& \ell_{21}& \ell_{42}\cr
e_3	& 1	& 1& \ell_{27}& \ell_{31}& 0& \ell_{35}& \ell_{36}& \ell_{37}& 0& \ell_{10}& \ell_{16}& \ell_{22}& \ell_{43}\cr
e_4	& 1	& 1& \ell_{28}& \ell_{32}& \ell_{35}& 0& \ell_{38}& \ell_{39}& 0& \ell_{11}& \ell_{17}& \ell_{23}&  \ell_{44}\cr
e_5	& 1	& 1& \ell_{29}& \ell_{33}& \ell_{36}& \ell_{38}& 0& \ell_{40}& \ell_8& \ell_{12}& \ell_{18}& \ell_{24}& \ell_{45}\cr
e_6	& 1	& 1& \ell_{30}& \ell_{34}& \ell_{37}& \ell_{39}& \ell_{40}& 0& \ell_9& \ell_{13}& \ell_{19}& \ell_{25}& \ell_{46}\cr
b_1	& 0	& 0& \ell_{6}& \ell_{7}& 0& 0& \ell_{8}& \ell_{9}& 1& 0& \ell_{2}& \ell_{4}& \ell_{47}\cr
b_2	& 0	& 0& 0& 0& \ell_{10}& \ell_{11}& \ell_{12}& \ell_{13}& 0& 1& \ell_{3}& \ell_{5}& \ell_{48}\cr
z_1	& 1	& 1& \ell_{14}& \ell_{15}& \ell_{16}& \ell_{17}& \ell_{18}& \ell_{19}& \ell_{2}& \ell_{3}& 1& \ell_{1}& \ell_{49}\cr
z_2	& 1	& 1& \ell_{20}& \ell_{21}& \ell_{22}& \ell_{23}& \ell_{24}& \ell_{25}&\ell_{4}& \ell_{5}& \ell_1& 1& \ell_{50}\cr
\alpha & 1	& 1& \ell_{41}& \ell_{42}& \ell_{43}& \ell_{44}& \ell_{45}& \ell_{46}& \ell_{47}+1& \ell_{48}+1& \ell_{49}+1& \ell_{50}& \ell_{51}\cr
  }
  \mod 2\,.
\end{align}
As every $c_{ij}$ set corresponds in principle to a different model,  simple counting gives a huge number of $2^{51}\sim 2.3\times 10^{15}$ distinct models in this class. 
Thus a comprehensive scan of even this restricted class of models would take 3000 years on a single core CPU.

Nonetheless, the models share some common attributes. First the gauge group  $G=SU(4)\times{SU(2)}_L\times{SU(2)}_R\times U(1)^3 \times SO(4)^2\times SO(8)$. Second the untwisted sector matter states comprise
six $\left(\mb{6},\mb{1},\mb{1}\right)$ representations and a number of PS singlets. The 
twisted sector states that transform nontrivially under the PS gauge symmetry include
the ``spinorial'' states  $\left(\mb{4},\mb{2},\mb{1}\right)$, $\left(\mb{4},\mb{1},\mb{2}\right)$, $\left(\omb{4},\mb{2},\mb{1}\right)$,  $\left(\omb{4},\mb{1},\mb{2}\right)$ and the 
``vectorial'' states $\left(\mb{1},\mb{2},\mb{2}\right)$, 
$\left(\mb{6},\mb{1},\mb{1}\right)$. The former arise from the sectors $b^I_{pqrs}\left(+S\right),$ $I=1,2,3$ and the latter from $x+b^I_{pqrs}\left(+S\right),$ $I=1,2,3$,
where $b^1_{pqrs}=b^1+p\,e_3+q\,e_4+r\,e_5+s\,e_6$, $b^2_{pqrs}=b^2+p\,e_1+q\,e_2+r\,e_5+s\,e_6$, $b^3_{pqrs}=x+b^1+b^2+p\,e_1+q\,e_2+r\,e_3+s\,e_4$, $p,q,r,s\in\{0,1\}$, and $x=\mathds{1}+S+\sum_{i=1}^6e_i+\sum_{k=1}^2z_k$. 

Additional exotic states transforming as 
$\left(\mb{4},\mb{1},\mb{1}\right)$, $\left(\omb{4},\mb{1},\mb{1}\right)$ $\left(\mb{1},\mb{2},\mb{1}\right)$ and $\left(\mb{1},\mb{1},\mb{2}\right)$ under the observable PS gauge group may also arise from the twisted sectors $b^I+\alpha \left(+z_1\right)\left(+x\right)\left(+S\right)$, $I=1,2,3$. We denote by $n_e$ the number of these states. They carry fractional charges and in particular they include SM singlets and doublets with $\pm\frac{1}{2}$ electric charge. The appearance of these states is generic in these vacua
\cite{gfm}.  However, as shown in \cite{pshm} the class of models under consideration  includes ``exophobic''  vacua  where  all exotic fractionally charge states receive string scale masses.


Selecting amongst this huge number of vacua requires first the computation of the spectrum and second the introduction of a set of phenomenological criteria.  As illustrated in \cite{cformalism} 
we can derive general analytic formulae regarding the main  characteristics 
of models in this set
in terms of the GGSO phases, $\ell_i,i=1,\dots,51$. These formulae involving ranks of binary matrices depending on $\ell_i$ are too lengthy to include here. However, they can be easily incorporated in 
a computer code.  The model selection criteria can be either related to the spectrum or
to the couplings of the effective low energy theory. 
The latter are harder to implement so we will restrict to the existence of the top quark mass coupling. As demonstrated recently \cite{top} this requirement can be expressed explicitly in terms of constraints on the GGSO phases,
\begin{align}
\label{topcup}
\ell_i=0,i=2,\dots,7\  ,\ \ell_{10}=\ell_{11}=\ell_{47}=0\  ,\ \ell_{48}=1\ ,\ \ell_{8}=\ell_{12}\  ,\ \ell_{9}=\ell_{13}\ .
\end{align}
Let us summarise therefore the possible selection criteria. We may choose to impose:\vspace{0.3cm}\\
\begin{tabular}{l}
 (a) 3 complete family generations, $n_g=3$ \\
 (b) Existence of PS breaking Higgs, $k_R\ge1$ \\
 (c) Existence of SM Higgs doublets, $n_h\ge1$ \\
 (d) Absence of exotic fractional charge states, $n_e=0$\\ 
 (e) Existence of top Yukawa coupling as in eq.(\ref{topcup}).\\ 
\end{tabular}\vspace{0.3cm}\\
 A more stringent test would be to insist on minimality by imposing $k_R= n_h=1$. \\

\section{GAs in the fermionic string landscape}

 \subsection{Introductory remarks}

Let us now see how a GA performs in the search for viable models. First we make some general remarks. 
When it comes to string phenomenology any fitness landscape is composed not of continuous functions but
of physical properties such as supersymmetry, number of generations,
Yukawa couplings and so forth. Nevertheless the question of whether
the fitness landscape defined in terms of such observables has structure
remains crucial, and one of the purposes of testing GA's is therefore
to address this issue. 

To be more specific, suppose that one constructs a GA to converge on models 
with three generations. To do this
would require a fitness function perhaps of the form $f(n_{g})=e^{-(n_{g}-3)^{2}}$;
that is models are weighted with a Gaussian around the desired value.
Clearly the population will
coalesce around $n_{g}=2,3$ or $4$ rather than $n_{g}=10$ but as
emphasised in the Introduction, for just one parameter, this way of
selecting vacua is not obviously much more beneficial than a random scan;
the GA procedure only really gives
advantage once the search is multi-modal with many
different criteria coming into play.

We can at this point mention 
the two studies of the string landscape using GA's in ref.\cite{Blaback:2013ht} (which along with ref.\cite{david} 
are the only other works on GA's in the context of string phenomenology of which we are aware).
These searches are also multi-modal, but mainly because there are many different fluxes contributing to the 
same single phenomenological trait, namely the vacuum energy. Thus multi-modality can result from 
either an already existing many-to-one mapping of genotype to phenotype, or from a many-to-many mapping when 
several phenomenological requirements are introduced. The situation here, where we are searching for discrete 
phenomenological constraints, is really of the latter kind; 
selecting on the single trait of three generations would not be multi-modal enough to make GA's beneficial, as we shall see.

For example suppose that we now add the requirement that only one
coupling (i.e. the top Yukawa) is large. It could be that the schemata
that favour a large top Yukawa overlap in the genotype with those
that give three generations. If they do not then it is equivalent
to adding a second dimension in our introductory example, and the
two phenomenological requirements can be decomposed into orthogonal
constraints on the definitions of the models. If there are enough
orthogonal constraints chopping up the genotype, then suitable models
can conceivably be constructed by hand.
Conversely, if the schemata do overlap then the fitness landscape
is much more complicated. And it could also be that top Yukawas and
three generations are incompatible. In this way as more search criteria
are applied, one expects more structure.

A more formal discussion of structure requires some measure of distance in the 
search space. One method for defining closeness in search spaces is the ``Hamming
distance'', based on the number of ``moves'' that have to be made
on one genotype in order to make it identical to the other. 
(More generally one can allow for insertions and deletions -- known as the Levenshtein distance.
In the models we will consider, where we are adjusting only a fixed number of GGSO phases, the Hamming distance is 
obviously the appropriate measure.) 
This is a useful concept when deciding if a GA will be effective on a certain problem, and 
it is important to realise that what one normally considers to be close phenomenologically may be 
far apart as far as a GA is concerned.

For instance suppose we have constructed an $SU(5)$ model with a
certain set of basis vectors. The Standard Model is then achieved
by introducing a projection onto the spectrum with an additional basis
vector that encodes gauge symmetry breaking. In terms of the Hamming
distance the two models are clearly far apart -- the data describing
the Standard Model differs by an entire basis vector from that of
the GUT model. However physically speaking the models are closely
related: in some more complete formalism the symmetry breaking can
probably be encoded by simply turning on a modulus. Conversely models
that we would usually regard as being very different (such as $SU(5)$
and $SU(7)$) may differ by only a few entries in their defining vectors.
Moreover it is known that identical models can be described by different
sets of basis vectors. For example often new basis vectors
project out states from the spectrum but at the same time give rise
to new sectors in which they reappear.
 
Clearly one expects that if selecting preferred models gets one closer to 
some nearby solution, then the GA will work efficiently. This has been formalised in the {\em fitness distance correlation} (FDC), that is the correlation 
between the \emph{fitness} and the \emph{Hamming distance to the nearest solution} \cite{jonesforrest}. It has been  argued that generally the better this correlation is, the more effective 
GA's are at finding a solution (with problems fall into one of three classes depending on the FDC, \emph{misleading, difficult}, and \emph{straightforward}).  
In the above example, it could be that $SU(5)$ GUT models and the Standard Model are incompatible for a GA because they 
are far apart in terms of Hamming (Levenshtein) distance; any fitness function that favoured both models would necessarily have a poor FDC.

Unfortunately in the present context, and in many contexts
where GA's are advantageous, the FDC will be extremely difficult to establish directly. 
This is because in order
to do so one has to first locate the position of \emph{every} solution 
so that one can compute the Hamming distance to the nearest
one. But if one can deterministically find every solution then the
problem is probably not
interesting for GA's anyway. In computational complexity terms, establishing the correlation
is clearly at least as hard as the well-known NP-hard ``closest string'' problem. 

\subsection{The GA analysis}

In summary it seems that, in many situations, examining the convergence properties of GA's
may actually be the best tool we have to study the structure of the landscape. Let us therefore now turn to the Pati-Salam models 
outlined in the previous section and describe the GA search. 

We will consider three classes of solutions, that are increasingly phenomenologically
viable, and hence increasingly ``statistically difficult''. In the first
class we ask for three complete generations of SM multiplets ($n_g=3$) together with
at least one Pati-Salam and SM Higgs ($k_R\geq 1$, $n_h\geq 1$): we call these the SM-complete
models and a random search shows their number to be roughly one in
$10^{4}$. In the second class of models we also insist that there
are no exotics ($k_R\geq 1$, $n_h\geq 1$, $n_e=0$), the exophobic models. Roughly one in $2.5\times10^{6}$
models are in this class. Finally in the third class we also ask for
a non-zero top Yukawa coupling ($k_R\geq 1$, $n_h\geq 1$, $n_e=0$, $Y_t\neq 0$). 
From the GA perspective this last requirement is interesting because it is a direct constraint on the schemata: namely eq.(\ref{topcup}) is equivalent to the condition that one of the following schemata is present: 
\begin{align}
\label{topcup2}
& S_{\rm top}=*~(0)^6 ~0 ~0 ~0~0~ 0 ~0~ (*)^{33}~0~1~(*)^3 \nonumber \\
& S_{\rm top}=*~(0)^6 ~0 ~1 ~0~0~ 0 ~1~ (*)^{33}~0~1~(*)^3 \nonumber \\ 
& S_{\rm top}=*~(0)^6 ~1 ~0 ~0~0~ 1 ~0~ (*)^{33}~0~1~(*)^3 \nonumber \\
& S_{\rm top}=*~(0)^6 ~1 ~1 ~0~0~ 1 ~1~ (*)^{33}~0~1~(*)^3 \, .
\end{align}

Our main tool for examining the convergence properties of the GA is the average number of string models that one has to construct before finding 
a solution, which we refer to as the ``call count''. Obviously in a completely random scan this would be the same as the ``statistical difficulty'', i.e. 
$10^4$, $2.5\times 10^6$, $10^{10}$ respectively. Note that this is also the expected call-count for the GA if one dials the mutation rate up to 100\%. 

The GA is organised as follows. The genotype is taken to have length 51 and consist of the $\ell_i$ binary numbers defining the GGSO phases.  
The optimal population was found to be 50 members. (Convergence rate by generation stays roughly constant if the population is increased, 
so it merely reduces the efficiency.) Assigning each criterion a desired value $\mu_i$ and actual value $\nu_i$, the fitness function was taken to be 
of the form $f(\mu_i,\nu_i)=10-\gamma_i (\mu_i-\nu_i)^2$ where $\gamma_i$ is a factor to give each criterion equal weight. Note that the $\mu_i$ are 
in this context integers (e.g. $n_g$ for the generation number, or $0$ or $1$ for the existence or otherwise of a top Yukawa). In fact the precise form of $f$ 
(the $\gamma_i$ for example)  does not significantly alter the convergence unless extreme values are taken. 

Searches were performed for different background mutation rates, $\mu_{\rm b'grd}$. The technique for tackling stagnation was as described above; if the maximum fitness was unchanged for more than 8 generations then a 
large mutation of $25\times  \mu_{\rm b'grd}$ was carried out over the whole population\footnote{Again these are numbers that have to be optimised empirically. 
If stagnation is flagged too early for example then the algorithm is unable to converge on a solution.}. 
Finally after each solution was found the entire population was scrambled and the 
search restarted to find the next solution. 

The call count using this procedure can be drastically less than that of a random scan for the statistically more 
difficult searches. This is shown in figure \ref{fig1} where we show the call count for the three different classes of solution. For the least constrained models 
we see that the GA is not much of an improvement over a random scan, as expected. However for those searches with a statistical difficulty of $2.5\times 10^6$ and $10^{10}$ 
we see that the GA is many orders of magnitude more efficient. Moreover the efficiency is extremely sensitive to the background mutation rate; 
it is optimised at around $\mu_{\rm b'grd}= 0.75-1\%$. That is the mutation probability per bit is optimally 0.0075-0.01. This is a clear sign that the GA is working as expected. The efficiency drops dramatically when the mutation is turned off completely (when the population is unable to discover new favourable schemata and/or stagnates) and also when the mutation is dialled up and the search becomes effectively randomised. It is close to but slightly below the rate $1/l \approx 0.02$ which is often claimed to be the optimal rate \cite{ga-books}.

\begin{figure}
\noindent \centering{}\includegraphics{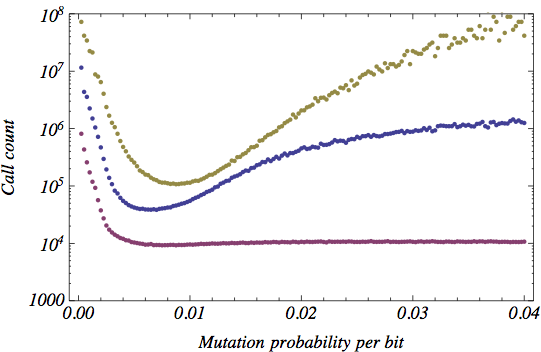}\caption{\label{fig2} \emph{Call count the for three different classes of solutions with increasing
search difficulty. (i.e. the mean number of models one has to construct
before finding a solution.) Bottom/purple: solutions with three generations
and Higgses for the Standard Model and Pati-Salam sectors. Middle/blue:
solutions with three generations, Higgses for the Standard Model and
Pati-Salam sectors, and in addition no exotics. Top/yellow: solutions
with three generations, Higgses for the Standard Model and Pati-Salam
sectors, no exotics and a top-Yukawa. The search difficulties are
respectively one in $10^{4}$, one in $2.5\times10^{6}$, one in $10^{10}$. }}
\end{figure}

Although there are only three points of reference it is worth noting that the minimal call count 
appears to be increasing roughly as the log of the statistical difficulty and slower than a power law; 
empirically we find call-count $\approx  7\times 10^3 \log({\rm diff}/4\times 10^3)$.  It would be of 
interest to make this relationship more precise. 

There is one further probe of the structure we can make. Instead of completely scrambling the genotypes after a solution is discovered, one can instead 
perform the same mutation of $25 \times \mu_{\rm b'grd}$ that one does when the population stagnates. If this yields new solutions (i.e. the population 
should not simply revisit the same solution) at a faster rate, then 
this indicates that the solutions are ``clustered'' together (in terms of Hamming distance) rather than spread uniformly. This would certainly be expected if the system is 
modular with different non-overlapping schema governing different phenomenological traits. More generally it would imply that the solutions occupy a hypersurface in the search space. 

This is equivalent to ``seeding'' the starting value from a previously found solution. Generally such a procedure  is known to introduce positive and negative aspects. On the positive side the algorithm is more efficient, but on the negative side it can lead to premature convergence and may lead the algorithm to entirely miss islands of favourable solutions. 

With his caveat in mind, the results for the most phenomenologically complete solutions (with a $10^{10}$ search difficulty) are shown in figure \ref{fig2}. As can be seen the call count drops again (by a factor of 25) at the optimal mutation rate (although obviously it has to return to the same non-GA values as the mutation rate goes to 1 or 0). Although this may not seem dramatic, one should recall that when one completely randomises after finding each solution, this is the rate achieved with a much 
lower difficulty $10^4$. In accord with expectations, one finds that the new solutions are dominantly the same as the old ones with just one or two properties such as number of PS Higgses altered. 

\subsection{Possible improvements}

The analysis presented above is sufficiently optimised for the constrained set of models considered here. In the future, and especially to perform a more general analysis of string vacua where the genotypes defining a model might have say 500 loci, further optimisation may be useful to consider\footnote{An aspect that is already optimal is that, at least in the fermionic formulation, string models are already largely in a binary format.}. 

One crucial aspect that we have so far not discussed in depth is the selection process. For this analysis we used simple roulette wheel selection, but it is known that some forms of selection may be more efficient when the population size becomes large. One can quantify the difference by measuring the ``takeover time'', namely the time for the best string to take over the whole population. In roulette wheel selection this typically increases as $p\log p$ with population, $p$, whereas with tournament selection (in which members of the old population compete with each other for breeding) it is known to increase only as $\log p$. Secondly we have here assumed a constant mutation rate, except where there was premature convergence at which point it was temporarily increased. It may be interesting to instead vary the mutation rate more continuously, and possibly to introduce a crowding penalty. 

\begin{figure}
\noindent \centering{}\includegraphics{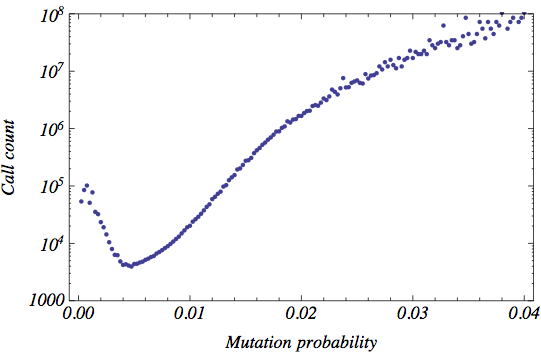}\caption{\label{fig1} \emph{Call count when the population is mutated away from the previous solution 
rather than being completely randomised, for the solutions with three generations, Higgses for the Standard Model and Pati-Salam
sectors, no exotics and a top-Yukawa. The search difficulty is one in $10^{10}$.}}
\end{figure}

\section{Concluding remarks}

We see this paper as a step towards more widespread use of genetic algorithms and other heuristic methods in the search for viable 
string vacua. We have restricted the discussion to Pati-Salam models in the Free Fermionic Formulation of 
heterotic strings in order to be able to compare GA's with simple randomised scans, as the latter can be performed very quickly in this class of models. 

We found that GA's are many orders of magnitude more efficient at discovering solutions with desirable phenomenological characteristics. It is in the nature of heuristic 
search methods that rigorous expressions for the expected search time are hard to come by. Nonetheless we find  
weak evidence that at least in this small subset of models the search time for GA's to find a solution increases only as the logarithm of the 
statistical difficulty. It would be interesting to study this relation in more detail; if it holds it would indicate that these and similar heuristic methods would be 
crucial for searches with more refined criteria than the ones considered here. 

It would also of course be interesting to now extend the analysis to more general configurations that include the possibility of different 
boundary vectors as well as generalised GSO phases, and also to consider other kinds of compactification. These are much more 
time consuming to construct and would be 
a real testing ground for GA's; the search space is many orders of magnitude larger than the relatively small one  of $2^{51}$ discussed here, and performing any kind of meaningful randomised search is not feasible.

\subsection*{}
{\bf {Acknowledgements:}} SAA would like to thank Jeppe Andersen, David Grellscheid and Tuomas Hapola for
 help with Grid computing and Keith Dienes and David Grellscheid for discussions. We would like to thank CERN Theory for hospitality while
this work was instigated. This research has been co-financed by the European Union (European
Social Fund-ESF) and Greek national funds through the
Operational Program "Education and Lifelong Learning", of the National
Strategic Reference Framework (NSRF) (Thales Research Funding Program) investing in knowledge society through the European Social Fund.

\end{document}